# Lab on a Chip

## PAPER

# On-demand contact line pinning during droplet evaporation

Wei Wang,§ Qi Wang,§ Kaidi Zhang, Xubo Wang, Antoine Riaud* and Jia Zhou*

Depending on the contact line motion, colloid-rich drolets evaporation can leave a ring-like or a spot-like residue. Herein, we determine this outcome by controlling the contact line motion using coplanar direct current electrowetting-on-dielectrics (DC-EWOD). Combined with theoretical calculations of the droplet shape and its evaporation rate, the time-dependent actuation voltage is first derived from experiments and simulations. Thanks to the additional control over the contact angle, the contact line can be maintained in pinned state even on surfaces that exhibit little contact angle hysteresis such as homogenous flat Teflon coatings. In the absence of EWOD control, polystyrene particles and *Escherichia coli* suspended in the droplet formed a dot-like pattern at the center of the initial contact, whereas application of the mechanism resulted in ring-like patterns of a controllable radius. Unlike chemically or structurally patterned substrates, the contact line could recover its mobility at any preset time before reaching the control limit, which is useful to accurately and consistently fabricate self-assembled nanostructures of desired patterns on different surfaces.

## Introduction

During the evaporation of droplets containing a colloidal suspension or other nonvolatile solutes such as bacteria, [1, 2] viruses [3] and proteins, [4, 5] a ring-like deposit often forms along the liquid-substrate contact line. This phenomenon, known as the coffee ring effect, [6] has led to numerous applications, including inkjet printing, [7] DNA deposition, [8] protein microarrays, [9, 10] cell patterning, [11] and cost-efficient biological assays [12, 13, 14]

This ring stain is due to the pinning of the interface [6]. Indeed, as the droplet evaporates, the mass loss at the droplet edge is compensated by a net radial flow from the droplet centre towards the edge. This flow may carry particles that accumulate at the edge as the droplet evaporates, thus forming the ring. When the pinning is supressed, the droplet leaves a spot-like residue. [1, 15-18] Hence, in order to improve the reliability of the aforementioned applications, controlling the contact line pinning is of primary importance.

The simplest models of droplet evaporation either assume a constant contact radius (CCR) where the contact line is pinned and the contact angle changes, or a constant contact angle (CCA) where the contact line is mobile. [1, 15-18] However, the evaporation of droplets on solid substrates tends to follow a stick-slip mode (CCR and CCA modes dominate alternatively) or a mixed mode (the contact angle and contact line change simultaneously). [19-23] During this quasi-equilibrium process, the pinning of the contact line is mainly attributed to the hysteresis effect. [24] When the deformation of droplets overcomes the hysteresis barrier, the contact line recovers its mobility and the accumulation of solutes at the boundary ends.

Various passive methods have been proposed to delay the transition from CCR to CCA, including self-pinning, [25] micro-patterns, [26-29] surface heterogeneities, [23, 27] additives, [30] non-spherical particles, [31] nanoparticles on the boundary, [32] and particulate-assisted droplet spreading. [33] Nevertheless, the application scenarios of these technologies are limited due to their strong dependence on droplet composition and surface modification. The irreversible nature of surface modification also complicates integration with other functional components. Indeed, current technologies do not allow pinning a droplet depending on an external sensor readout. Moreover, the pinning duration is not adjustable in real-time. With an active pinning technology, the ring formation could be made less sensitive to the droplet composition if the droplet by releasing the contact line once a thick enough ring had formed. Furthermore, releasing the droplet early would also increase the system throughput by requiring less time for evaporation. Hence, active methods to control the coffee-ring formation are highly desirable.

Recently, electrostatic control of the droplet contact angle, namely electrowetting-on-dielectric (EWOD), has served for on-demand colloidal self-assembly and patterning. [34-36] The system does not require complex modification of the substrate, and can even be promoted to non-hydrophobic non-flat surfaces. [37-39] Notably, microscopic boundary vibrations induced by alternating current EWOD (AC-EWOD) can eliminate the hysteresis effect of droplets and suppress the occurrence of the ring stain. [36, 40, 41] This observation suggests that the direct current electrowetting-on-dielectric (DC-EWOD) could dynamically control the droplet boundary and tune the deposition pattern. However, EWOD was introduced in previous works only as a mean of liquid transportation or vortex generation. Its capability of dynamically tuning the droplet morphology has long been neglected in evaporation-related processes.

*ASIC and System State Key Laboratory, School of Microelectronics, Fudan University, Shanghai 200433, China. E-mail: antoine_riaud@fudan.edu.cn, jia.zhou@fudan.edu.cn*

§ These authors contribute equally to this work.

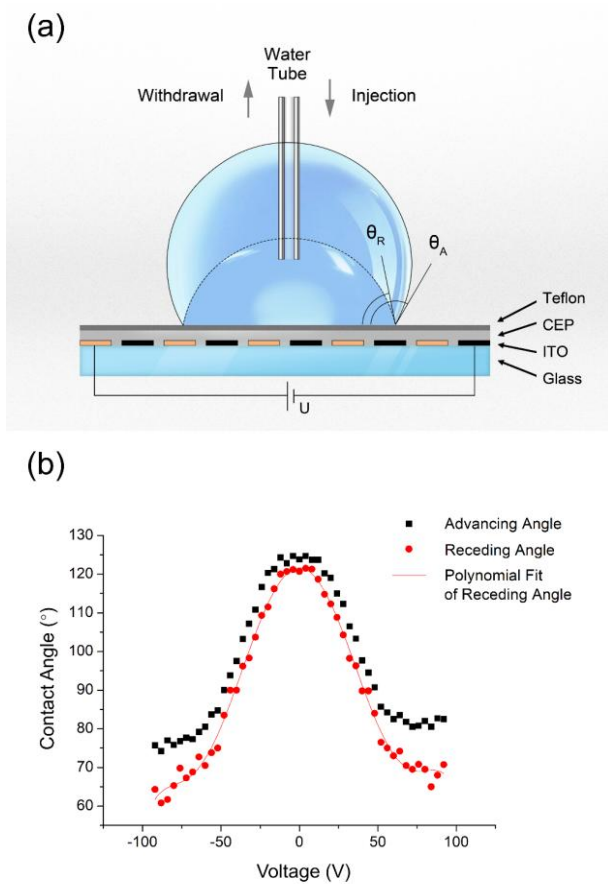

**Fig. 1** Characterization of coplanar DC-EWOD structures. (a) Experimental setup. Sessile droplets were deposited and drawn under DC-EWOD at a constant rate of 0.1 μL/s via a steel tubing. Upon inflating the droplet, the contact radius increases at the advancing contact angle $\theta_A$. When reversing the pumping direction, the contact radius decreases at the receding contact angle $\theta_R$. (b) Contact angles under DC-EWOD. The contact angles are averaged over 5 independent measurements of left and right edges (with a standard deviation less than 4°).

In this paper, the actuation voltage of a coplanar DC-EWOD device was continuously adjusted to pin the contact line of an evaporating droplet to a predetermined position. In consequence, the CCR duration of the evaporating droplet lasts 7 times longer. After the evaporation, ring patterns of a predetermined contact radius were formed on hydrophobic surfaces. This outcome contrasts sharply with the much smaller dot-like deposits of an uncontrollable radius formed without applied voltage. We then illustrate the biological applications of this system by concentrating microparticles and bacteria into a ring.

## Setup of the coplanar DC-EWOD

The coplanar DC-EWOD system is sketched as Fig. 1a. Straight interdigitated electrodes (50 μm of finger width and spacing, with a 2 cm aperture) were patterned on a 130 nm-thick indium-tin-oxide coated glass (purchased from Wesley Technology Co., Ltd) by standard photolithography and wet etching. Cyanoethyl pullulan (purchased from Shin-Etsu Chemical Co., Ltd.) and Teflon® AF2400 (purchased from DuPont) were successively spin-coated on the surface as a dielectric layer (thickness of 410nm) and a hydrophobic layer (thickness of 60 nm), respectively. Deionized water was deposited and drawn back at a constant rate (0.1 μL/s) via a steel tubing with direct current (DC) supplies conducted exclusively to the electrodes. Side view images of the droplet were captured to determine the contact angle using a commercially available goniometer (DSA30, KRÜSS, Germany). [42]

## Results and Discussion

**Performance of the coplanar DC-EWOD**

For each electrical potential $U$, though a specific equilibrium contact angle $\theta$ is predicted by the Lippmann-Young equation, [24] some hysteresis is observed. This appears in Fig. 1b as a gap between the contact angles of inflating and shrinking droplets. An additional complication is that droplets also exhibit a saturation phenomenon for high voltages ($U_{sat} = 50V$). Since there is still no well-verified quantitative theories to describe these deviations, we fitted our experimental results with a polynomial to obtain an accurate $\theta = \theta(U)$ relation for subsequent use,. Given the shrinking droplet volume in the evaporation process, we consider only the receding contact angles presented in Fig. 1b. A sixth-degree polynomial was chosen because fitting curves of lower degrees show an obvious discrepancy with experimental results while higher degree polynomials tend to over-fit the data by behaving more like wavy lines. The fitting algorithm yielded:

$$\frac{\theta(U)}{\theta(U=0)} = 1 - 0.503\left(\frac{U}{U_{sat}}\right)^2 + 0.191\left(\frac{U}{U_{sat}}\right)^4 - 0.024\left(\frac{U}{U_{sat}}\right)^6, \#(1)$$

with $\theta(U=0) = 121.6°$ the contact angle of deionized water on Teflon in the absence of applied voltage. Due to the symmetry between negative and positive voltages, the odd-degree coefficients were vanishingly small and have been omitted. This fitting method can similarly apply to devices with other structures.

According to this experimental characterization, the droplet equilibrium contact angle can be dynamically adjusted to remain

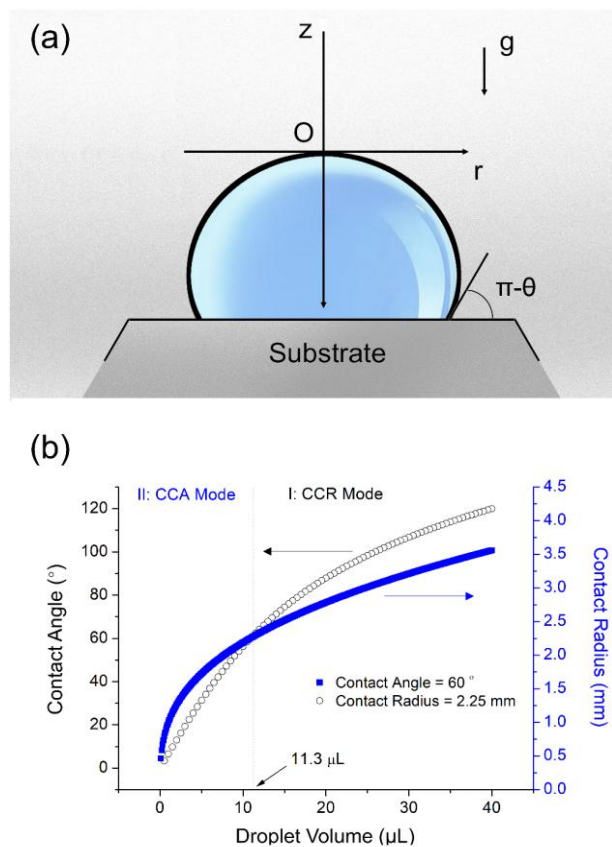

**Geometrical estimation of the droplet dimension**

To geometrically link the contact angle to the volume of droplets with a pinned contact line, and thus choose an appropriate actuation voltage, we need to compute the dimension of droplets resting over flat surfaces. As shown in Fig. 2a, the origin of the coordinates is located at the top of the droplet residing on a flat substrate. The $z$-axis is parallel to the gravity $g$. To ensure accuracy, we numerically solved the differential Young-Laplace equation for the axisymmetric shape of droplets in the presence of gravity:

$$-\frac{r''}{(1+r'^2)^{\frac{3}{2}}} + \frac{1}{r(1+r'^2)^{\frac{1}{2}}} = \frac{2}{b} + \frac{g\Delta\rho}{\gamma_{lg}}z, \#(2)$$

where the liquid/gas density difference and interfacial tension are presented by $\Delta\rho$ and $\gamma_{lg}$, respectively and the superimposed prime denotes the derivative with respect to $z$. In these calculations, the height of the droplet $h$ and the radius of curvature at the origin point $b$ were not known a priori. For each candidate value of $b$, the constant $h$ was obtained by incrementing $z$ until the droplet profile reached the desired radius or contact angle. Thus, each value of $b$ was associated with a specific droplet volume.

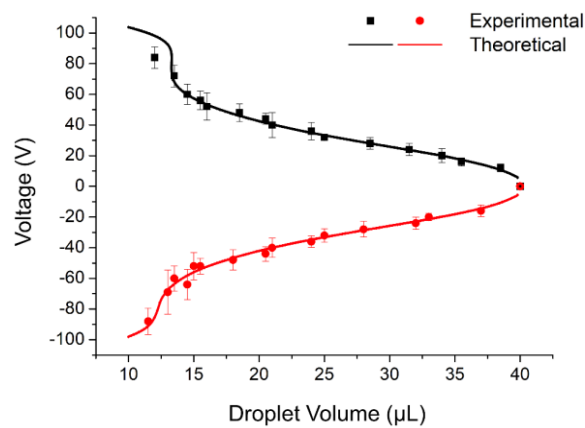

**Fig. 2** Calculated droplet dimension as a function of the droplet volume. (a) Equilibrium shape of an aqueous droplet residing on a flat substrate in the gravity field. The origin of the coordinate is located at the top of the droplet. The contact angle is $\theta$. (b) Simulated contact radius and contact angle versus droplet volume. The contact radius of a 40 µL droplet with a 121.6° contact angle is 2.25 mm. In this work, the CCR mode dominates until the contact angle reaches the control limit of the EWOD mechanism at 60°. This process is labeled as stage I: CCR Mode. According to the numerical model, the remaining volume is 11.3 µL at this transition threshold. The contact line recovers its mobility afterward and slides in the CCA mode towards the end of the evaporation. This is labelled as stage II: CCA Mode.

**Fig. 3** Voltage required to pin droplets of various volumes to a predetermined position. The contact radius of the circular contact area is fixed at 2.25 mm. The contact angle under no actuation voltage is 121.6°. The black and red colors present data for positive and negative voltages, respectively. Numerically calculated lines fit well with the experimental dots.

The calculated contact angle is shown in Fig. 2b as a function of the droplet volume $V$ for a fixed contact radius 2.25 mm, which is denoted by $\theta = \theta(V)$. This contact radius is calculated for a 40 µL droplet with a 121.6° contact angle. Given the ability of coplanar DC-EWOD to control the contact angle in this work (as shown in Fig. 1b), the relation for the contact radius versus the droplet volume below the recessing contact angle, thus effectively pinning the contact line of an evolving droplet geometry. This pinning can be maintained for a broad range of contact angle until it drops below 60°. This is in stark contrast with previous EWOD experiments where the receding contact angle could not be adjusted depending on the droplet state. To predict this time-dependent actuation voltage, a semi-empirical approach was developed based on experimental calibration. It combines (i) a relation between the contact angle and actuation voltage (as shown in Fig. 1b), (ii) a geometrical estimation of the droplet contact radius versus the volume, and (iii) a theoretical expression for the evaporation rate of droplets. The derivation process is detailed hereafter.

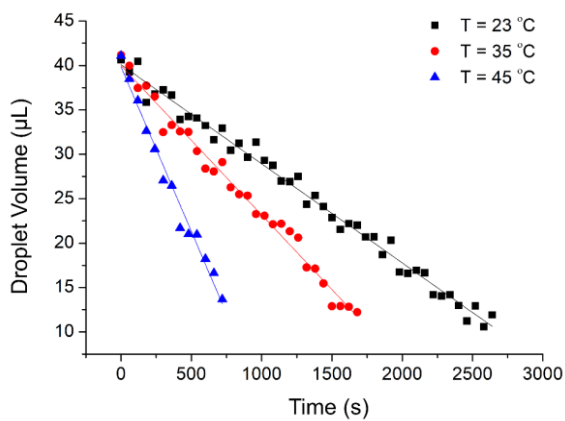

**Fig. 4** Evaporation of droplets on heated substrates. The weight of droplets was measured by an analytical balance and then converted to the volume. The evaporation rate is constant for each test at the CCR mode and increases when the temperature increases. The lines are guides for the eye.

is also presented in Fig. 2b at a constant contact angle of 60° (the lower limit where droplet contact line will escape the control mechanism and recover its mobility). According to our calculations, the volume of a droplet with a 2.25 mm contact radius and a 60° contact angle is 11.3 μL. The two curves have adjusted to visually intersect at the threshold where the mode transition happens. It is worth noting that the contact line slides at the CCA mode on our Teflon-coated substrates, but may end up in either CCA, stick-slip or mixed mode depending on the properties of other potential substrates.

The tuning mechanism ensures that the contact line of an evaporating droplet pins to its initial position until the contact angle decreases below the control limit. The key factor in this process is to dynamically adjust the voltage following the volume reduction of the droplet. Otherwise, the contact line would slide to counteract the deformation. In this respect, Fig. 3 is derived by combining the geometrically calculated $\theta = \theta(V)$ relation with the fitted $\theta = \theta(U)$ polynomial. It presents the voltage supply required to pin droplets of different volumes to a predetermined position (i.e. a circular contact area of 2.25 mm radius here). For convenience, the relation is denoted as $U = U(V)$.

**Evaporation of sessile droplets on heated substrates**

Eventually, to dynamically accommodate the volume reduction with a time-dependent actuation voltage, it is necessary to obtain the evaporation rate of droplets. According to the experimental results depicted in Fig. 4, the droplet volume decreases linearly with the evaporation time in the CCR mode, in agreement with previous works.[19, 22, 43] The CCR mode ends when the contact angle of droplets decreases below 60° at the threshold volume of 11.3 μL.

With saturated vapour concentrations at the substrate temperature $T_s$ and the ambient temperature $T_a$ denoted as $c_s(T_s)$ and $c_s(T_a)$, respectively, Popov[44] and Gelderblom et al.[45] described the evaporation rate of droplets from a theoretical perspective:

$$\frac{dM}{dt} = -\pi RD[c_s(T_s) - Hc_s(T_a)]f(\theta), \#(3)$$

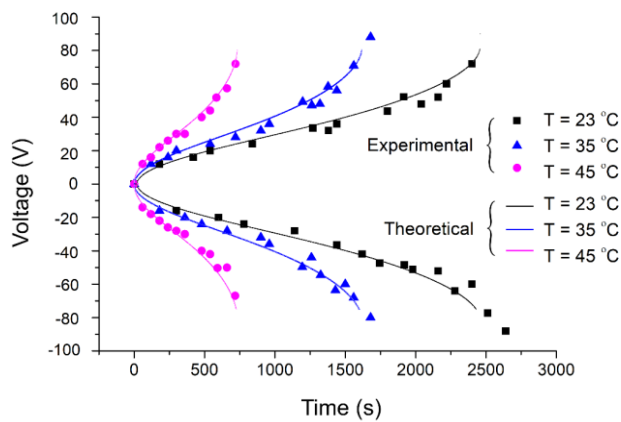

**Fig. 5** Time-dependent actuation voltage as a function of the evaporation time at various temperatures. The contact angle is dynamically tuned by smoothly rising the actuation voltage to avoid the boundary sliding due to the volume shrinkage. The solid line presents the theoretical estimation of our model. The initial volume of the droplet is 40 μL with a contact angle of 121.6°.

where $M$ is the droplet mass, $t$ is the evaporation time, $H$ is the relative humidity in the ambient air, $D$ is the diffusion constant of water vapour in air and $f$ reads:

$$f(\theta) = \frac{\sin\theta}{1+\cos\theta} + 4\int_0^\infty \frac{1+\cosh 2\theta\tau}{\sinh 2\pi\tau}\tanh[(\pi-\theta)\tau]\,d\tau. \#(4)$$

Once the contact angle $\theta$ is known, the evaporation rate $dM/dt$ is derived and then integrated to get the mass loss of droplets versus the evaporation time. Given that the mass and the corresponding volume of a droplet could be easily converted via its density, we achieve the calibration of the actuation voltage versus evaporation time by substituting Eq. (3) in $U = U(V)$.

**Time-dependent actuation voltage**

Following the derivation above, the time-dependent actuation voltage is obtained by combining Eq. (1-4). The numerical model was verified experimentally for various substrate temperatures as shown in Fig. 5. Regardless of the temperature, the mechanism significantly delays the occurrence of the CCR-to-CCA mode transition. For Teflon-coated hydrophobic surfaces, the threshold contact angle decreases from the initial 121.6 ° to a much lower value of 60 °. According to our calculations for the droplet dimension, the corresponding unpinning threshold volume decreases from 35 μL (no regulation) down to 11.3 μL (with regulation). This results in a 7-fold extension of the evaporation time under CCR mode as detailed in Table 1:

Table 1: CCR mode duration with and without EWOD control, at various temperatures.

| Temperature (°C) | CCR mode duration without EWOD control (s) | CCR mode duration with EWOD control (s) |
|---|---|---|
| 23 | 360 | 2640 |
| 35 | 230 | 1740 |
| 45 | 115 | 960 |

Theoretical estimations (solid lines) agree well with the experimental data. In general, the droplet evaporates faster for a higher temperature. This requires the actuation voltage to grow more rapidly to accommodate the volume loss. Moreover, the contact line could recover its mobility at any predetermined time before reaching the control limit simply by connecting the electrodes to the ground. It makes the mechanism an on-demand method.

**Optimization of dried spots for *Escherichia Coli* biosensors**

*Escherichia coli* (*E. coli*) is key model bacteria and has been connected with diarrhoea, kidney failure, haemorrhagic colitis, haemolytic uremic syndrome. Among the variety of label-free biosensor sensors, surface acoustic wave devices stand out due to their sensitivity to surface perturbation and mass loading of the targeted specimen. However, as the targeted specimen is normally loaded by the dried spot of droplets on the device surface, irregularities of deposit patterns generate a significant impact on the final readout. [46]

Herein, we demonstrate the capacities of our technology by controlled evaporation of sessile droplets and the accurate formation of a coffee-ring. 40 μL water droplets containing a suspension of polystyrene particles (600 nm diameter, 0.05% volume fraction) or *E. coli* (2000 bacteria per droplet) were evaporated on a 23 °C substrate. Between each experiment, the surface was rinsed by deionized water to eliminate minute amounts of particle sediments that formed outside of the main deposit.

In a control experiment, droplets were naturally dried on a Teflon-coated substrate. As a consequence, typical dried deposits for a droplet-on-hydrophobic system formed as shown in Fig. 6a and Fig. 6b, yielding dot-like patterns of 0.78 mm±0.02 mm and 0.95 mm±0.01 mm diameters for polystyrene particles and *E. coli*, respectively. The low hysteresis of the hydrophobic surface ensures that the contact line of the droplet slides at a high contact angle until the concentration in the residual droplet becomes large enough to trap the contact line. That makes the naturally dried patterns vary for droplets of different initial specimen concentrations, which adds an unpredictable variable in the test. Besides, the deposit pattern is affected by numerous factors including solute concentration, surfactant concentration, ionic strength, and particle properties. A comprehensive study of these parameters was conducted by Deegan. [47]

When the contact-line control mechanism is activated, the evaporation duration in CCR mode is primarily dependent on EWOD ability to tune the contact angle. The resulting ring stain of evaporation in CCR mode after 2640 s evaporation is shown in Fig. 6c and Fig. 6d. Consistent with our calculations, the radius of ring-like deposits averaged over 20 tests is 2.35 mm±0.09 mm and 2.84 mm±0.06 mm for polystyrene particles and *E. coli*, respectively. Hence, the ring radius is relatively insensitive to the droplet content. Compared to spot-like deposits, the large radius of the *E. coli* ring contributes to its adhesion to the substrate. The width of the *E. coli* ring is 118 μm±13 μm, which is uniform and controllable with our mechanism. Although here single droplets already contained enough particles to form a clear ring, this system could be integrated with other EWOD systems to bring additional droplets and produce a more abundant deposit and thus, decrease detection thresholds.

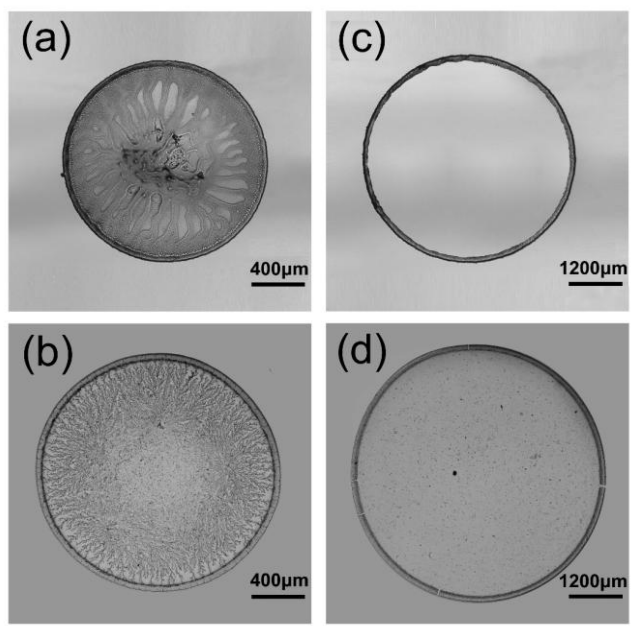

**Fig. 6** Deposits of dried droplets containing a suspension of (a, c) polystyrene particles and (b, d) *Escherichia coli* while the mechanism was introduced (c, d) or not (a, b). The residual liquid after evaporation was removed and the background was rinsed then by deionized water.

## Conclusions

Contact line pinning technologies are limited in terms of throughput, robustness and integration. We have developed a framework to pin the boundary of a droplet on demand using coplanar DC-EWOD, by dynamically tuning the droplet morphology to accommodate the volume shrinkage in the evaporation process. To this end, we measured the contact angle of droplets versus the actuation voltage and then related it to the droplet volume by numerical calculations. Feeding this electro-mechanical model with the evaporation rate of droplets on a heated substrate, we eventually derived the time-dependent actuation voltage. Thanks to this strategy, we observed a 7-fold increase in the CCR duration and the controllable formation of ring stains on hydrophobic surfaces. Our device is relatively easy to fabricate and operate. As the voltage can be flexibly programmed, the proposed system offers an effective way for on-demand patterning. Furthermore, it provides an optimal approach to accurately load the targeted specimen on the device surface, which can also be conveniently integrated with other digital microfluidic units for multifunctional platforms.

## Conflicts of interest

There are no conflicts to declare.

## Acknowledgements

This work was supported by the National Science Foundation of China with Grant No. 61704169, the National Natural Science Foundation of China with Grant No. 61874033, the Natural Science Foundation of Shanghai Municipal with Grant No. 18ZR1402600, and the State Key Lab of ASIC and System, Fudan University with Grant No. 2018MS003. The work was also supported by the China Scholarship Council (CSC) with File No. 201706100083.

## References

1   W. Sempels, R. De Dier, H. Mizuno, J. Hofkens, J. Vermant, *Nat. Commun.*, 2013, **4**, 1757.
2   A. K. Thokchom, R. Swaminathan, A. Singh, *Langmuir*, 2014, **30**, 12144-12153.
3   R. Gebhardt, J. –M. Teulon, J. –L. Pellequer, M. Burghammer, J. –P. Colletier, C. Riekel, *Soft Matter*, 2014, **10**, 5458-5462.
4   H. M. Gorr, J. M. Zueger, J. A. Barnard, *Langmuir*, 2012, **28**, 4039-4042.
5   H. M. Gorr, J. M. Zueger, D. R. McAdams, J. A. Barnard, *Colloids Surf. B*, 2013, **103**, 59-66.
6   R. D. Deegan, O. Bakajin, T. F. Dupont, G. Huber, S. R. Nagel, T. A. Witten, *Nature*, 1997, **389**, 827-829.
7   J. Park, J. Moon, *Langmuir*, 2006, **22**, 3506-3513.
8   R. Blossey, A. Bosio, *Langmuir*, 2002, **18**, 2952-2954.
9   Y. Deng, X. –Y. Zhu, T. Kienlen, A. Guo, *J. Am. Chem. Soc.*, 2006, **128**, 2768-2769.
10  L. H. Mujawar, J. G. M. Kuerten, D. P. Siregar, A. van Amerongen, W. Norde, *RSC Adv.*, 2014, **4**, 19380-19388.
11  E. A. Roth, T. Xu, M. Das, C. Gregory, J. J. Hickman, T. Boland, *Biomaterials*, 2004, **25**, 3707-3715.
12  D. Zhang, B. Gao, Y. Chen, H. Liu, *Lab Chip*, 2018, **18**, 271-275.
13  J. T. Wen, C. –M. Ho, P. B. Lillehoj, *Langmuir*, 2013, **29**, 8440-8446.
14  C. P. Gulka, J. D. Swartz, J. R. Trantum, K. M. Davis, C. M. Peak, A. J. Denton, F. R. Haselton, D. W. Wright, *ACS Appl. Mater. Interfaces*, 2014, **6**, 6257-6263.
15  N. Murisic, L. Kondic, *J. Fluid Mech.*, 2011, **679**, 219–246.
16  H. Hu, R. G. Larson, *Langmuir*, 2005, **21**, 3963–3971.
17  F. Girard, M. Antoni, K. Sefiane, *Langmuir*, 2008, **24**, 9207–9210.
18  R.G. Picknett, R. Bexon, *J. Colloid Interface Sci.*, 1977, **61**, 336–350.
19  S.Y. Misyura, *Appl. Therm. Eng.*, 2017, **113**, 472–480.
20  H. –Z. Yu, D. M. Soolaman, A. W. Rowe, J. T. Banks, *ChemPhysChem*, 2004, **5**, 1035–1038.

21  X. Fang, M. Pimentel, J. Sokolov, M. Rafailovich, *Langmuir*, 2010, **26**, 7682–7685.
22  R. Mollaret, K. Sefiane, J. R. E. Christy, D. Veyret, *Chem. Eng. Res. Des.*, 2004, **82**, 471–480.
23  Y. –F. Li, Y. –J. Sheng, H. –K. Tsao, *Langmuir*, 2013, **29**, 7802-7811.
24  U. –C. Yi, C. –J. Kim, *J. Micromech. Microeng.*, 2006, **16**, 2053-2059.
25  B. M. Weon, J. H. Je, *Phys. Rev. Lett.*, 2013, **110**, 028303.
26  Y. V. Kalinin, V. Berejnov, R. E. Thorne, *Langmuir*, 2009, **25**, 5391-5397.
27  S. Wang, K. Liu, X. Yao, L. Jiang, *Chem. Rev.*, 2015, **115**, 8230−8293.
28  J. Xia, M. Su, *Lab Chip*, 2017, **17**, 3234-3239.
29  B. Balu, A. D. Berry, D. W. Hess, V. Breedveld, *Lab Chip*, 2009, **9**, 3066-3075.
30  S. F. Shimobayashi, M. Tsudome, T. Kurimura, *Sci. Rep.*, 2018, **8**, 17769.
31  P. J. Yunker, M. A. Lohr, T. Still, A. Borodin, D. J. Durian, A. G. Yodh, *Phys. Rev. Lett.*, 2013, **110**, 035501.
32  Y. Li, H. Wu, F. Wang, *Langmuir*, 2016, **32**, 12676-12685.
33  D. Mampallil, M. Sharma, A. Sen, S. Sinha, *Phys. Rev. E*, 2018, **98**, 043107.
34  O. Kudina, B. Eral, F. Mugele, *Anal. Chem.*, 2016, **88**, 4669-4675.
35  J. Zhang, M. K. Borg, K. Ritos, J. M. Reese, *Langmuir*, 2016, **32**, 1542-1549.
36  H. B. Eral, D. Mampallil Augustine, M. H. G. Duits, Frieder Mugele, *Soft Matter*, 2011, **7**, 4954-4958.
37  M. Abdelgawad, S. L. S. Freire, H. Yang, A. R. Wheeler, *Lab Chip*, 2008, **8**, 672-677.
38  J. –Y. Cheng, L. –C. Hsiung, *Biomed. Microdevices*, 2004, **6**, 341-347.
39  F. Lapierre, M. Jonsson-Niedziolka, Y. Coffinier, R. Boukherroub, V. Thomy, *Microfluid. Nanofluid.*, 2013, **15**, 327-336.
40  F. Li, F. Mugele, *Appl. Phys. Lett.*, 2008, **92**, 244108.
41  F. Mugele, A. Staicu, R. Bakker, D. van den Ende, *Lab Chip*, 2011, **11**, 2011-2016.
42  J. Chen, Y. Yu, K. Zhang, C. Wu, A. Q. Liu, J. Zhou, *Sens. Actuators, B*, 2014, **199**, 183-189.
43  X. Jiang, L. Tian, X. Liu, T. Li, *Colloids Surf., A*, 2018, **545**, 31-38.
44  Y. O. Popov, *Phys. Rev. E*, 2005, **71**, 036313.
45  H. Gelderblom, Á. G. Marń, H. Nair, A. van Houselt, L. Lefferts, J. H. Snoeijer, D. Lohse, *Phys. Rev. E*, 2011, **83**, 026306.
46  S. T. Ten, U. Hashim, S. C. B. Gopinath, W. W. Liu, K. L. Foo, S. T. Sam, S. F. A. Rahman, C. H. Voon, A. N. Nordin, *Biosens. Bioelectron*, 2017, **93**, 146-154.
47  R. D. Deegan, *Phys. Rev. E*, 2000, **61**, 475-485.